\documentclass[twocolumn,letter]{jpsj2} 
\def\degangle{\kern-.2em\r{}} 

\title{
Synthesis, Characterization, and Magnetic Properties of $\gamma$-Na$_{x}$CoO$_{2}$ ($0.70\leq x\leq 0.84$)
}

\author{
Hiroya \textsc{Sakurai}$^{1}$\thanks{E-mail address: sakurai.hiroya@nims.go.jp}, Satoshi \textsc{Takenouchi}$^{2}$, Naohito \textsc{Tsujii}$^{3}$, and Eiji \textsc{Takayama-Muromachi}$^{1}$
}

\inst{
$^{1}$Superconducting Materials Center, National Institute for Materials Science (SMC/NIMS), 1-1 Namiki, Tsukuba, Ibaraki 305-0044, Japan.\\
$^{2}$Research Promotion Division, National Institute for Materials Science, 1-1 Namiki, Tsukuba, Ibaraki 305-0044, Japan.\\
$^{3}$NanoMaterials Laboratory, National Institute for Materials Science (NML/NIMS), 1-2-1 Sengen, Tsukuba, Ibaraki 305-0044, Japan.\\
}

\abst{
Powder Na$_{x}$CoO$_{2}$ ($0.70\leq x\leq 0.84$) samples were synthesized and characterized carefully by X-ray diffraction analysis, inductive-coupled plasma atomic emission spectroscopy, and redox titration. It was proved that $\gamma$-Na$_{x}$CoO$_{2}$ is formed only in the narrow range of $0.70\leq x\leq 0.78$. Nevertheless, the magnetic properties depend strongly on $x$. We found, for the first time, two characteristic features in the magnetic susceptibility of Na$_{0.78}$CoO$_{2}$, a sharp peak at $T_{\mbox{p}}=16$ K and an anomaly at $T_{\mbox{k}}=9$ K, as well as the transition at $T_{\mbox{c}}=22$ K and the broad maximum at $T_{\mbox{m}}=50$ K which had already been reported. A type of weak ferromagnetic transition seems to occur at $T_{\mbox{k}}$. The transition at $T_{\mbox{c}}$, which is believed to be caused by spin density wave formation, was observed clearly for $x\geq 0.74$ with constant $T_{\mbox{c}}$ and $T_{\mbox{p}}$ independent of $x$. On the other hand, ferromagnetic moment varies systematically depending on $x$. These facts suggest the occurrence of a phase separation at the microscopic level, such as the separation into Na-rich and Na-poor domains due to the segregation of Na ions. The magnetic phase diagram and transition mechanism proposed previously should be reconsidered.
}

\kword{Na$_{x}$CoO$_{2}$, ICP, redox titration, solid solution, magnetic susceptibility, SDW, ferromagnetic, phase separation}

\begin{document}
\maketitle

The strongly correlated electron system (SCES) is one of the most interesting research subjects for condensed matter physicists because of its excellent characteristics such as unconventional superconductivity, metal-insulator transition, and giant magnetoresistance. Na$_{x}$CoO$_{2}$ has attracted attention as an SCES since the discovery of high thermoelectrical performance\cite{NaCo2O4N1,NaCo2O4N2}. The number of research studies on this compound has increased recently because of the discovery of superconductivity in its analogue, Na$_{x}$CoO$_{2}$$\cdot$yH$_{2}$O\cite{NCO}. $\gamma$-Na$_{x}$CoO$_{2}$ has a layered structure in which oxygen planes with a triangular lattice are stacked in an AABB sequence. Co and Na ions occupy the octahedral and prism sites, respectively. The CoO$_{6}$ octahedra share their edges forming CoO$_{2}$ layers with Co ions on a triangular lattice.

Many experimental studies on $\gamma$-Na$_{x}$CoO$_{2}$ have been performed, and the results have not always been consistent. Na$_{0.75}$CoO$_{2}$ shows a transition to the spin density wave (SDW) state at $T_{\mbox{c}}=22$ K\cite{MotohashiSDW}, which was confirmed by muon spin rotation/relaxation ($\mu$SR) measurement\cite{SugiyamaSDW1}. Although SDW usually causes an increase in electric resistivity, it drops just below $T_{\mbox{c}}$ in the present compound.  Motohashi $et$ $al.$ have proposed a mechanism that the resistivity drop is due to the elimination of one of two Fermi surfaces by the SDW formation\cite{MotohashiSDW}, based on the band calculation of Na$_{0.5}$CoO$_{2}$\cite{Singh}. This anomaly at $T_{\mbox{c}}$ has been reported by other groups\cite{Prabhakaran,Bayrakci} for samples with slightly different $x$ values. However, neutron scattering measurements have shown that the dominant magnetic fluctuation is not antiferromagnetic, but ferromagnetic\cite{Boothroyd}. Furthermore, it has been suggested that this compound seems to be an itinerant electron system\cite{Boothroyd}. On the other hand, it has been recently reported that no such transition is seen in the magnetic susceptibility of the $x=0.75$ sample, which shows a broad maximum at 14 K\cite{Miyoshi}, or Curie-Weiss behavior (with a tiny kink at approximately 30 K probably caused by the secondary phase of Co$_{3}$O$_{4}$)\cite{Carretta}.

The solid solution range of Na content has been reported to be narrow between $x$=0.65 and 0.78\cite{xRange}. For a lower $x$ of 0.5-0.7, however, almost all samples contained Co$_{3}$O$_{4}$ as an impurity phase, which showed a kink or a peak in the magnetic susceptibility at approximately 30-35 K. According to nuclear magnetic resonance (NMR) measurements, the resultant phase was always Na$_{0.7}$CoO$_{2}$ when the nominal $x$ was lower than 0.7 \cite{Alloul}. These results indicate that the solid-solution range terminates at $x\simeq 0.7$. In addition, Na$_{0.7}$CoO$_{2}$ may have a charge-ordered state of magnetic Co$^{4+}$ ($3d^{5}$) and nonmagnetic Co$^{3+}$ ($3d^{6}$) at a low temperature as suggested by NMR\cite{Alloul,Ray,Gavilano}, although it shows metallic behavior below 300 K. It is suggested that the magnetic Co$^{4+}$ ions induce Curie-Weiss behavior in the magnetic susceptibility.

It should be noted that recent research of the Na$_{x}$CoO$_{2}$ system is focused on the two end members of the solid solution with $x\simeq 0.7$ and $\simeq 0.78$.  These members seem to have quite different physical properties with apparent itinerant-electron behavior for $x=0.75$, and with localized-electron behavior for $x=0.7$. On the other hand, a dome-shaped magnetic phase diagram with $x$ as a variable has been proposed\cite{SugiyamaSDW2,SugiyamaSDW3} in spite of no experimental data between $x=0.65$ and 0.75. In the present study, we synthesized Na$_{x}$CoO$_{2}$ samples and investigated their magnetic properties systematically. 

The powder samples of Na$_{x}$CoO$_{2}$ ($x=0.70$, 0.72, 0.74, 0.76, 0.78, 0.80, 0.82, and 0.84) were synthesized by conventional solid state reaction from the stoichiometric mixture of Na$_{2}$CO$_{3}$ (99.99\%) and Co$_{3}$O$_{4}$ (99.9\%), which were dried at 300\degC\ before use. The mixture was cold-pressed into a pellet, and in order to avoid possible vaporization of Na, the surface-to-volume ratio of the pellet was reduced and the heating process was carried out carefully. More concretely, the amount of each mixture was about 12 $g$ and the mixture was pressed into several pellets with a diameter of 15 $mm$. The pellets were placed in a dense aluminous crucible and heated in flowing oxygen gas three times at $T_{\mbox{syn}}=800$, 850 and 950\degC\ with intermediate grindings. The increase rate of temperature was 5\degC/min up to 100\degC\ below $T_{\mbox{syn}}$ and was reduced to 20\degC/h for the last 100\degC\ increase to minimize Na vaporization. The sample was maintained at $T_{\mbox{syn}}$ for 6 h and cooled in the furnace. In the final heat treatment at 950\degC, cooling rate was controlled; it took 2 h from 950 to 850\degC, 1 h from 850 to 750\degC, 6 h at 750\degC\ for annealing, and 12 h from 750 to 350\degC. We tried to synthesize Na$_{0.67}$CoO$_{2}$ but a single-phase sample could not be obtained even after heating several times. The grain sizes of the present samples seemed to vary less compared with those made by the rapid heating technique\cite{RHtech}. The obtained samples were kept in an evacuated desiccator to prevent absorption of water in air.

The samples were characterized by powder X-ray diffraction (XRD) analysis, inductive-coupled plasma atomic emission spectroscopy (ICP-AES) and redox titration. XRD analysis was carried out using a Bragg-Brentano-type diffractometer (RINT2200HF, Rigaku) with Cu $K_{\alpha}$ radiation. Lattice constants were calculated from several peaks in the range of 5\degangle$\leq 2\theta \leq 60$\degangle. ICP-AES was carried out by dissolving the sample in hydrochloric acid to determine the ratio of Na to Co, $x$. In the redox titration, each sample was dissolved in hydrochloric acid with an excess amount of (COONa)$_{2}$ as a reducing agent, and residual (COONa)$_{2}$ was titrated using a standard aqueous solution of KMnO$_{4}$. The Co valence, $s$, was estimated from the result of the titration using the weight ratio determined by ICP-AES.

The magnetic properties of the powder samples were measured using a commercial magnetometer with a superconducting quantum interference device (MPMS-XL, Quantum Design). Before each measurement under zero-field cooling (ZFC) conditions, the magnetic field was reset to 0 Oe at $T=100$ K or 300 K. In this paper, we simply define magnetic susceptibility as $\chi =M/H$ ($M$: magnetization, and $H$: applied magnetic field).

\begin{figure}
\begin{center}
\includegraphics[width=7cm,keepaspectratio]{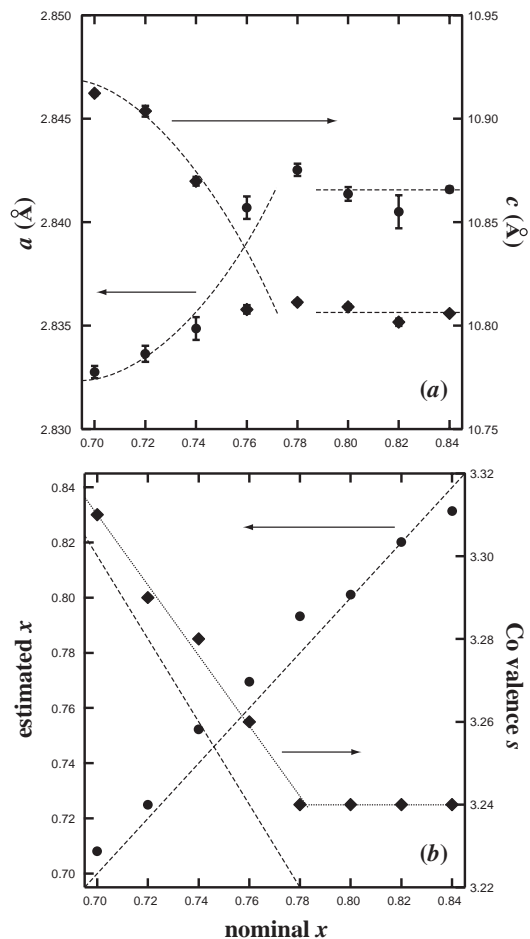}
\end{center}
\caption{
(a) Lattice constants, $a$ (the circles) and $c$ (the squares), of Na$_{x}$CoO$_{2}$. The broken lines are visual guides. (b) Ratio of Na to Co, $x$, estimated by ICP-AES (the circles) and the formal valence of Co, $s$, estimated by the redox titration (the squares). The broken lines are the ideal $x$ and $s$ expected from the nominal $x$. The dotted lines are the fitting results (see the text).
}
\label{LP}
\end{figure}

All XRD peaks were assignable to $\gamma$-Na$_{x}$CoO$_{2}$ for every sample. However, this does not necessarily mean only a single phase was present, because a possible secondary phase such as Na$_{2}$CO$_{3}$ or NaOH is hard to detect by XRD analysis. Indeed, in Fig. \ref{LP}(a), the lengths of the $a$- and $c$-axes vary with $x$ below $x\sim 0.78$ but are constant for $x$ above 0.78 , strongly suggesting the higher solution limit of $x$=0.78.

The results of ICP-AES and the redox titration are shown in Fig. \ref{LP}(b). The ratio of Na to Co, $x$, determined by ICP-AES is in good agreement with the nominal value even for a higher $x$ range. This means that Na was not vaporized in our synthetic process.  Ideally, the Co valence, $s$, should obey the line, $s=4-x$, but experimental data are placed on the line, $s=3.9-0.85x$, below $x=0.78$. Above $x=0.78$, the Co valence has the constant value of 3.24. The two lines have a difference of $\sim$0.01 in $s$ which seems to be beyond the accuracy of our titration experiment and may be due to a certain systematic error.  Nevertheless, we believe that the almost constant Co valences above $x=0.78$ reflect the solution limit of $x$=0.78. The synthesis experiment is consistent with the solid solution range of 0.70 - 0.78 in $x$ and this is supported by the magnetic measurements (see below).

The $T$-dependences of $\chi$ of Na$_{0.78}$CoO$_{2}$ are shown in Fig. \ref{chi078}. There are four characteristic features: (1)the sharp transition at $T_{\mbox{c}}=22$ K, (2) a fairly sharp peak at $T_{\mbox{p}}=16$ K of FC curves under $H\leq 100$ Oe, (3) a broad maximum at $T_{\mbox{m}}=50$ K, and (4) an anomaly at approximately $T_{\mbox{k}}=9$ K which is seen as an upturn under a low magnetic field while as a dull downward kink under a high magnetic field of 10 kOe $\leq H\leq 30$ kOe.

\begin{figure}
\caption{
(a) Temperature dependence of magnetic susceptibility of Na$_{0.78}$CoO$_{2}$ measured under various magnetic fields. The inset shows the $M$-$H$ curve at 1.8 K. The solid line is $M=-3.18\times 10^{-4}+1.55\times 10^{-7}H$. (b) Expanded graph of (a). The $\chi$-$T$ curves are off-set to be distinguished.
}
\label{chi078}
\end{figure}

The transition at $T_{\mbox{c}}$ has been reported by Motohashi $et$ $al$\cite{MotohashiSDW} to be due to SDW formation. $T_{\mbox{c}}$ is completely independent of $H$ as seen Fig. \ref{chi078}. This transition and the hysteresis between ZFC and FC curves are also observed in other samples with $x\geq 0.74$, but is most prominent for $x=0.78$, less pronounced with decreasing $x$, and almost invisible for $x\leq 0.72$ as seen in Fig. \ref{chiAll}(b). However, $T_{\mbox{c}}$ is independent of $x$, which means that the magnetic phase diagram does not have a dome shape as suggested by Sugiyama $et$ $al$\cite{SugiyamaSDW2,SugiyamaSDW3}.

\begin{figure}
\caption{
Magnetic susceptibilities of Na$_{x}$CoO$_{2}$ measured under 10 kOe (a) and 100 Oe (b). The closed and open circles represent the susceptibilities and reciprocal susceptibilities, respectively.
}
\label{chiAll}
\end{figure}

\begin{figure}
\begin{center}
\includegraphics[width=7cm,keepaspectratio]{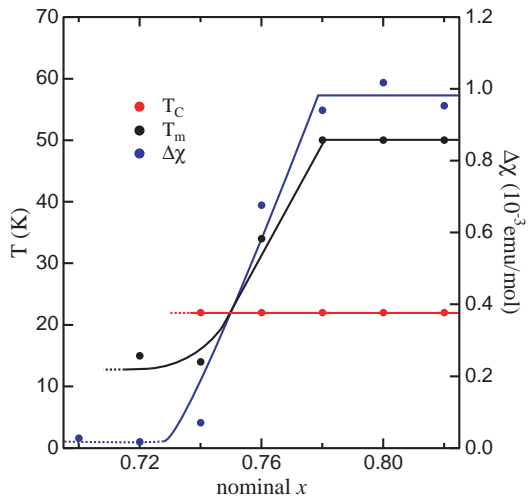}
\end{center}
\caption{
Characteristic parameters, $T_{\mbox{c}}$ (red), $T_{\mbox{m}}$ (black), and $\Delta\chi$ (blue). The lines are visual guides.
} 
\label{Para}
\end{figure}

$T_{\mbox{p}}$ is also independent of $x$ as seen in Fig. \ref{chiAll}(b). On the other hand, the difference in $\chi$ under 100 Oe between ZFC and FC data at $T_{\mbox{p}}$, $\Delta\chi$, increases systematically with increasing $x$ as seen in Fig. \ref{Para}. These strongly suggest that the phase responsible for the transition in question does not change but only its fraction varies with $x$. Thus, we considered a phase separation to explain these phenomena. A phase separation into Na-rich and Na-poor domains seems to be the most probable and the transition at $T_{\mbox{c}}$ takes place only in the former domain whose fraction increases with $x$. This idea is consistent with the specific heat measurement\cite{Cpandrho} and with the results of $\mu$SR which indicate that the volume fraction of magnetic domain is only $\sim$21\%\ in the case of $x=0.75$\cite{SugiyamaSDW1}. The transition has been observed in a single crystal\cite{Bayrakci} and the phase separation seems to occur at the microscopic level which cannot be detected by XRD analysis.

$T_{\mbox{m}}$ depends on $x$ as seen in Figs. \ref{chiAll}(a) and \ref{Para}. One may think that the broad maximum at $T_{\mbox{m}}$ is due to the SDW fluctuation, namely short-range or intralayer ordering. However, the $\chi$ of $x=0.74$ clearly shows $T_{\mbox{m}}<T_{\mbox{c}}$, which indicates that the broad maximum has no relation with the transition at $T_{\mbox{c}}$. The $x$ dependence suggests that this behavior is caused by domains that survive the phase separation. $T_{\mbox{m}}$ may correspond to a characteristic temperature below which a heavy Fermi-liquid state develops, as discussed previously\cite{Miyoshi}.

The data between 150 K and 320 K were fit using the Curie-Weiss law with a constant term. The Curie constants were approximately 0.16 emu$\cdot$K/Co mol with an effective moment of  $\simeq 1.1$ $\mu _{B}$. Assuming an ionic model with local moments caused by a low-spin configuration, only Co$^{4+}$ ions have the magnetic moment of $S=1/2$. Then, the Curie constant and effective moment of the Co$^{4+}$ ions are calculated to be 0.62 emu$\cdot$K/Co$^{4+}$ mol and 2.2 $\mu _{B}$, which almost agrees with the expected value of 1.7 $\mu _{B}$ in consideration of the deviation of the $g$-factor from 2. The Weiss temperature is approximately $-120$ K, from which one may assume antiferromagnetic interaction. However, ferromagnetic fluctuation has been observed by neutron inelastic scattering for Na$_{0.75}$CoO$_{2}$ and the compound is suggested to be an itinerant electron system\cite{Boothroyd}.  Band calculation has also pointed out its itinerant ferromagnetic tendency\cite{Singh}. Thus, the negative Weiss temperature could be due to the Curie-Weiss behavior of itinerant electrons with a small exchange enhancement factor\cite{Moriya}, namely this compound behaves like a nearly ferromagnetic metal, at least, at the high-temperature region. Many itinerant electron systems show Curie-Weiss behavior with Curie constants expected  from the local moments\cite{TiBe2,Pd,ZrZn2}.

The upturns of the ZFC curves under $H\leq 5$ kOe below $T_{\mbox{k}}$ appear to be due to a paramagnetic impurity. However, the upturns become less pronounced with increasing $H$, and become dull downward kinks under 10 kOe $\leq H\leq 30$ kOe (see Fig. \ref{chi078}(b)). This phenomenon cannot be explained by magnetic impurities. Thus, it seems that the anomaly is due to a weak ferromagnetic transition like that observed at $T_{\mbox{c}}$. Indeed, preliminary specific heat measurements show an anomaly at $T_{\mbox{k}}$ in the case of $x\geq 0.74$\cite{Cpandrho}. The anomaly is not seen for the sample with $x\leq 0.72$. On the other hand, for $H=50$ kOe and 70 kOe, the magnetic phase is probably different from that for $H\leq 30$ kOe; the $M$-$H$ curve at 1.8 K bends at approximately $H\simeq 40$ kOe as seen in the inset of Fig. \ref{chi078}(a).

There is no marked change in magnetic properties between $x=0.70$ and 0.78. This suggests the lack of change in the electronic state, although the models of itinerant electrons and local moments have been proposed for lower and higher $x$ values, respectively\cite{Boothroyd,Alloul,Ray,Gavilano}. On the other hand, the magnetic properties for $x\geq 0.80$ are the same as those for $x=0.78$, which is very consistent with the solid solution limit of $x=0.78$ determined in the synthesis experiments.

Finally, we will discuss the possibility of two types of Fermi surface with one being insensitive to carrier density. Although such a situation may explain the present magnetic phenomena without assuming phase separation, the presence of only one type of Fermi surface is very likely for $x>0.7$ according to the results of the band calculation of the $x=0.5$ phase\cite{Singh} and angle-resolved photoemission spectroscopy\cite{ARPES1,ARPES2}. Here, it is worth noting that the electric resistivity of the present system drops below $T_{\mbox{c}}$ and that the phenomenon has been explained assuming the two types of Fermi surfaces. It seems that another reason needs to be found for the resistivity drop. Moreover, the peak in the $\chi$-$T$ curve at $T_{\mbox{p}}$ is not caused by simple SDW formation. In order to elucidate the physical properties of the system in more detail, research to obtain microscopic information such as NMR is indispensable and is in progress.

In conclusion, we synthesized powder samples of Na$_{x}$CoO$_{2}$ ($0.70\leq x\leq 0.84$) and characterized them carefully by XRD analysis, ICP-AES, and redox titration. It was proved that $\gamma$-Na$_{x}$CoO$_{2}$ can be obtained for the narrow range of $0.70\leq x\leq 0.78$. Nevertheless, the magnetic properties of this system depend strongly on $x$. We have found, for the first time, two characteristic features of the magnetic susceptibility of Na$_{0.78}$CoO$_{2}$, the sharp peak at $T_{\mbox{p}}=16$ K and an anomaly at $T_{\mbox{k}}=9$ K, as well as a transition at $T_{\mbox{c}}=22$ K and a broad maximum at $T_{\mbox{m}}$ which were previously reported. A kind of weak ferromagnetic transition seems to take place at $T_{\mbox{k}}$. The transition at $T_{\mbox{c}}$, which is believed to be due to SDW formation, is clearly observed for $x\geq 0.74$ with fixed $T_{\mbox{c}}$ and $T_{\mbox{p}}$ independent of $x$. Ferromagnetic moment decreases systematically with decreasing $x$. These suggest a type of phase separation, such as that into Na-rich and Na-poor domains. The magnetic phase diagram and mechanism of the transition proposed thus far should be reconsidered.

\section*{Acknowledgements}
We would like to thank K. Takada, T. Sasaki, A. Tanaka, M. Kohno (NIMS), and K. Ishida (Kyoto University) for their fruitful discussion. This study was partially supported by Grants-in-Aid for Scientific Research (B) from the Japan Society for the Promotion of Science (16340111). One of the authors (H.S) is a research fellow of the Japan Society for the Promotion of Science.

\end{document}